\documentstyle[11pt]{article}
\def\etal{{\it et al.\ }}

\def\Lya{Ly $\alpha$}
\begin{document}
\title{The TAUVEX UV Imager}
\author{Noah Brosch \\ The Wise Observatory and the School of Physics 
and Astronomy \\ Tel Aviv University, Tel Aviv 69978, Israel}
\maketitle

In early 1988 the Israel Space Agency (ISA) solicited academic and 
commercial research and development groups in Israel to propose  
scientific payloads for a National Scientific Satellite (NSS). Among 
the many proposals submitted, that of Tel Aviv University, to orbit 
a number of small, wide-field telescopes to image astronomical objects 
in the ultraviolet (UV), was finally selected with the highest priority. 
This payload is referred to below as the {\bf T}el {\bf A}viv 
University {\bf UV E}xperiment (TAUVEX).

Since the early 1970's there has not been a space experiment able
to image a wide field with reasonable angular resulution in the UV, which
operated for more than about 10 days. The only full sky survey in the UV 
was conducted by the TD-1 satellite and resulted in a catalog containing 
31,215 sources measured in four spectral bands. Selected regions, to
deeper levels that the TD-1 survey, were observed by telescopes from balloons,
rockets, or dedicated satellites. The deepest such partial surveys 
by wide-field imagers are
by the FOCA balloon telescope and by the UIT Shuttle-borne instrument.

Observations in the UV region longward of Lyman $\alpha$ (\Lya \,)
up to the atmospheric 
transmission limit of $\sim$3000\AA\, take advantage of the reduced sky 
background. This is because of a fortuitous combination of zodiacal 
light decreasing shortward of $\sim$3000\AA\, and other backgrounds 
remaining low up to near the geocoronal \Lya. In this spectral region 
it is therefore possible to observe faint astronomical sources with a 
high signal-to-noise ratio with a modest telescope (O'Connell 1987).

The sources best studied with small aperture telescopes are QSOs 
and AGNs, that radiate significantly in the UV. Other sources of UV 
photons are hot stars of various types, the most interesting being 
white dwarfs and mixed-type binaries. Young, massive stars, that 
emit copious amounts of UV radiation and ionize the interstellar medium, 
are important in the context of star formation and evolution of galaxies; 
here the advantage of a wide-field imager is obvious. This has been 
demonstrated amply by the UIT instrument flown on the Space Shuttle
(Stecher \etal 1992). 
The obvious advantages of TAUVEX here are reduced sky background, the 
longer observing time per target, and the long duration mission.


In June 1991 it was 
proposed that TAUVEX be launched and operated from the Spectrum 
R\"{o}ntgen-Gamma (SRG) spacecraft as part of the SODART (Soviet-Danish 
R\"{o}ntgen Telescope) experiment. The SRG satellite will be launched 
in late 1997 by Russia into a high elliptical four-day orbit. SRG is the 
first of a series of space astronomical observatories being developed
under the sponsorship of the Russian Academy of Sciences with financial
support of the Russian Space Agency. For SRG, the scientific support
comes from the Space Research Institute (IKI) of the Russian Academy of
Sciences, and technical support is given by the Babakin Institute of
the Lavotchkin Association.

SODART (Schnopper 1994) consists of two X-ray imaging telescopes, 
each with four focal-plane 
instruments, to perform observations in the 0.2-20 keV band. It images
a one-degree field of view with arcmin resolution. TAUVEX will provide SODART 
with aspect reconstruction and will assist SRG in pointing and position
 keeping. In September 1991 the TAUVEX experiment was officially invited 
to join other instruments aboard the SRG spacecraft. ISA agreed in November 
1991 to provide SRG with the TAUVEX instrument. The official confirmation 
from the Russian side was received in December 1991.

The TAUVEX imagers will operate on the SRG platform alongside numerous 
X-ray and $\gamma$-ray experiments. This will be the first scientific
 mission providing simultaneous UV-X-$\gamma$ observations of celestial
 objects. The instruments on SRG include, apart from SODART, the JET-X 
 0.2-10 keV imager with 40' FOV and 
10-30" resolution, the MART 4-100 keV coded aperture
imager ($6^{\circ}$ FOV and 6' resolution), the two F-UVITA  
 700-1000\AA\, imagers ($1^{\circ}$ FOV and 10" 
resolution), the MOXE all-sky X-ray burst detector in the 3-12 keV band, and
the SPIN all-sky $\gamma$-ray burst detector in the 10keV-10MeV band
with $0^{\circ}$.5 optical localization.

TAUVEX will be bore-sighted with SODART, JET-X, MART and F-UVITA, and will 
obtain simultaneous imaging photometry of objects in the UV with three 
independent telescopes. A combination of various filters will accomodate 
wide, intermediate and narrow spectral bands. These will be selected to 
take maximal scientific advantage of the stability of SRG, the image 
quality of the optics (90\% of the energy in $\sim$8"), and long staring 
times at each SRG pointing. During a single pointing it will be possible 
to change filters, thus more than three UV bands can be used on one 
observation. 


The present design of TAUVEX includes three co-aligned 20 cm diameter 
telescopes in a linear array on the same mounting surface. Each telescope 
images  54' onto photon-counting position-sensitive 
detectors with wedge-and-strip anodes. Such detectors are space-qualified 
and have flown in a number of Space Astronomy missions. The TAUVEX detectors 
were developed by Delft Electronische Producten (Netherlands) to provide
high UV quantum efficiency at high count rates. Most systems within
TAUVEX are at least doubly-redundant. The choice of three telescopes
with identical optics and detectors adds an intrinsic degree of
redundancy. More safeties are designed into the software. Because of 
SRG telemetry constraints, TAUVEX has to accumulate an image on-board,
instead of transmitting time-tagged photons. The drift of the SRG
platform is compensated within the payload, by tracking onto a $\sim$bright 
(m$_{UV}<10.5$ mag) star in the field of view. The tracking corrections
are used to register the collected events and are supplied to the SRG
orientation and stabilization system.

The payload was 
designed and is assembled by El-Op Electro-Optics Industries, Ltd., of 
Rehovot, the top electro-optical manufacturer of Israel, with continuous
support and  
supervision of Tel Aviv University astronomers.
The development of TAUVEX follows a number of stages, in which the predicted 
behavior is verified by extensive tests. El-Op already produced a number 
of models of the experiment that were delivered to the Russian constructors 
of the spacecraft. The delivered models include a size mockup, a mass and 
center of gravity model for satellite vibration tests, and a thermal 
simulation model. The latter, in particular, is identical to the flight 
model except for its lack of electronics and working detectors. All
construction details and surface finishes were included, the telescopes 
have actual aluminized mirrors,  etc. 

The thermal model was tested at an ESA (European Space Agency) facility 
in Germany in late-January 1993 prior to its shippment to Russia. The test 
was a full space simulation, including Solar radiation, and the measured 
behavior verified the theoretical model developed at El-Op. The thermal
model has now been installed on a model of the entire spacecraft,
which will be submitted to a full environmental test, including shocks
and vibrations appropriate for the PROTON-2 launch.

In April 1993 
El-Op completed the engineering model of TAUVEX, which contains 
operational electronics. After testing, this model was shipped 
to the Russian Space Research Institute in Moscow, where it has been 
tested intensively. In particular, during 1996 the SRG instrument  
teams conducted a series of Complex Tests, in which instruments were
operated together, as if they were on-board the satellite. Two more such tests
are planned, until the engineering models of the instruments are delivered
early in 1997 to the Lavotchkin Industries to be integrated in a full
spacecraft engineering model. At present, the SRG  schedule calls for
a launch by  the end of 1997.


In parallel with tests in Russia, the TAUVEX models are passing their 
qualification in Israel. During the first half of 1996 the Qualification
Model (identical to the flight model) has been vibrated and submitted to
shocks stronger than expected during the SRG launch. In September 1996 we
expect to proceed with the thermal-vacuum qualification. This test, which
lasts more than one month, checks the behavior of the instrument at
extreme temperatures and in high vacuum conditions. During these tests,
we project various targets onto the telescopes' apertures with a high-precision
60 cm diameter collimator that allow us to fully illuminate one of the three
telescopes. We test for resolution, distortion, photometric integrity,
spectral response, etc.

While the QM is being paced through the qualification process, El-Op
continues building the flight model (FM). The optical module, containing
the three telescopes, has already been built and adjusted. The integration
of the detectors and electronics will follow immediately upon the
completion of the thermal tests. The FM will be submitted to a burn-in
process, a low-level qualification, followed by an extended calibration in the
thermal vacuum chamber of El-Op.

The timetable of the SRG project calls for the upper part of
the SODART telescopes to be integrated with the X-ray mirrors at IABG, near
M\"{u}nchen, in Germany. TAUVEX, which requires very clean assembly conditions
and is connected to a mounting plate on the side of the SODART telescopes,
will be integrated at IABG at the same time. Following the integration, 
the entire top part of the SRG spacecraft will be transported to Russia 
to be tested at Lavotchkin and integrated with the rest of the scientific 
payload.


The combination of long observing periods per source offered by of SRG 
(typically 4 hours, up to 72 hours), and a high orbit with low radiation 
and solar scattered background, 
implies that TAUVEX will be able to detect and measure star-like objects 
of $\sim$20 mag with S/N=10. This corresponds to V$\simeq$22.5 mag QSOs,
given typical UV-V colors of QSOs; at least 10 such objects are  
expected in every TAUVEX field-of-view. During the 3 year guaranteed life of 
SRG at least 30,000 QSOs will be  observed, if the targets
 will be different and at high galactic latitude. This is $\sim5\times$
 more QSOs than catalogued now. The multi-band observations, combined
with ground-based optical observations, allow the simple separation of
QSOs from foreground stars.

Diffuse objects, such as nearby large galaxies, will be measured to a
surface brightness of m$_{UV}\simeq$20 mag/$\Box"$.  A survey of the Local 
Group galaxies and  nearby clusters of galaxies, 
that cannot be conducted with the Hubble Space Telescope because of its 
narrow field-of-view, will be a high priority item in the target list 
of TAUVEX.
TAUVEX will detect hundreds of faint galaxies in each high latitude field. 
The large number of galaxies at faint UV magnitudes is indicated
 by balloon-borne observations of the Marseilles CNRS group
(Milliard \etal 1992). Recently it became clear that the UV-bright galaxies
observed by FOCA may be related to those responsible for the Butcher-Oemler
effect in clusters of galaxies. It is even 
possible that most faint, high latitude UV sources are galaxies.

Our prediction models indicate that each high-latitude field will 
contain similar numbers of galaxies and stars. Allowing for a 
reasonable fraction of low-{\bf b} fields, we estimate that TAUVEX
will observe $\sim10^6$ stars, mostly early type and WDs.
The data collection of TAUVEX will represent the deepest 
UV-magnitude-limited survey of a significant fraction of the sky. 

An additional major contribution of our experiment to astrophysics is 
the unique opportunity to study time-dependent phenomena in all energy 
ranges, from MeV in the $\gamma$-ray band to a few eV in the UV,
 together with the other scientific instruments
on board SRG. The combination of 
many telescopes observing the same celestial source in a number of spectral
 bands  offers unparalleled oportunities of scientific research.
 For the first time, it will be possible to study the physics of accretion 
disks around black holes and neutron stars, from the hard X-rays to near 
the optical region. Other subjects of study  include the inner regions of 
QSOs and AGNs, where the physics of the accretion phenomenon, probably 
powering all such sources, are best studied with simultaneous
multi-wavelength observations. 

In preparation for TAUVEX, the science team at Tel Aviv University is
collaborating with the Berkeley Space Astrophysics Group 
in analyzing the UV images of the FAUST Shuttle-borne imager.
The analysis is combined with gound-based observations from the
Wise Observatory, to enhance the identification possibilities. 
In parallel with the hardware development, the Tel Aviv team is studying
 the physics of UV space sources. A predictor model was developed to 
calculate the expected number of UV sources to any observation direction. 
The model tested well against the few existing data bases of UV sources. 
We are also predicting UV properties of {\it normal} sources 
from their known optical properties. This will allow us to detect
 extraordinary sources, through a comparison of their {\it predicted} 
and {\it measured} UV properties. Finally, we are creating at Tel Aviv 
University a large and unique data base of UV astronomy, by combining a 
number of existing data sets obtained by various space missions.

{\bf Acknowledgements:} I am grateful for support of the TAUVEX
project by the Ministry of Science and Arts, and by the Israel Academy of 
Sciences. UV astronomy at Tel Aviv University is supported by  
the Austrian Friends of Tel Aviv University. A long collaboration in 
the field of UV astronomy with Prof. S. Bowyer from UC Berkeley,
supported now by the US-Israel Binational Science Foundation, 
is appreciated. I thank all my colleagues of the TAUVEX
team for dedicated work during these years, and I am grateful to the
Korean Astronomical Observatory for inviting me to attend
this meeting.

{\bf Homepage:} Information on TAUVEX, with pictures, is available at:

 http://www.tau.ac.il/\~\,benny/TAUVEX/.

 \section* {References}
\begin{description}
\item    Milliard, B., Donas, J., Laget, M., Armand, C. and Vuillemin, A. 
1992 Astron. Astrophys. {\bf 257}, 24.

\item O'Connell, R.W. 1987 Astron. J. {\bf 94}, 876.

\item Schnopper, H.W. 1994 Proc. SPIE {\bf 2279}, 412.

\item Stecher, T.P \etal 1992 Astrophys. J. Lett. {\bf 395}, L1.

\end{description}

\end{document}